\documentclass[epsfig,11pt]{article}
\usepackage{epsfig}
\usepackage{graphicx}
\usepackage{array}
\usepackage{color}
\usepackage{bm}
\usepackage{bbm}
\usepackage{bm}
\usepackage{threeparttable}
\usepackage{adjustbox}
\usepackage[nottoc,notlof,notlot]{tocbibind} 
\usepackage[titles]{tocloft} 
\usepackage{booktabs}
\usepackage{amsmath}
\usepackage[justification=centering]{caption}
\usepackage[T1]{fontenc}
\usepackage{amsmath}
\usepackage{amssymb}

\newcommand{\beq}{\begin{equation}}   
\newcommand{\eeq}{\end{equation}}
\newcommand{\beqn}{\begin{eqnarray}}   
\newcommand{\eeqn}{\end{eqnarray}}

\newcommand{\ra}{\rightarrow}
\newcommand{\matel}[3]{\langle #1|#2|#3\rangle}
\newcommand{\gsim}{\lower.7ex\hbox{$
\;\stackrel{\textstyle>}{\sim}\;$}}
\newcommand{\lsim}{\lower.7ex\hbox{$
\;\stackrel{\textstyle<}{\sim}\;$}}
\begin{document}

\begin{flushright}
FTPI-MINN-24-25\\
UMN-TH-4405/24\\
\end{flushright}

\vspace{1mm}

\begin{center}

{\Large QCD Chemistry: Remarks on Diquarks}\footnote{Based on the talk given at the 27$^{\rm th}$ High-Energy Physics International Conference in Quantum Chromodynamis (QCD24), Montpellier, July  8-12, 2024.} 

\vspace{3mm}
M. SHIFMAN 

{\em {\large 
\vspace{1mm}

 William I. FineTheoretical Physics Institute, University of Minnesota,
Minneapolis, MN 55455}}
\end{center}

\vspace*{3mm}

\begin{center} 

{\large{\bf Abstract}}

\end{center}
{\small In connection with recent discoveries of heavy-quark containing exotic states publications discussing $Qq$ diquarks ($Q,q$ stand for a heavy and light quarks, respectively) proliferated in the literature. After a brief summary of the diquark concept I review various general reasons why the $Qq$ diquark (with sufficinetly heavy $Q$) does not exist. Then I argue (this is the focus of my talk)
that the most direct way to confirm non-existence of the $Qq$ diquarks is the study of pre-asymptotic corrections in the inclusive decays 
of  $Qqq$ baryons, e.g. $\Lambda_b$. Since the $c$ quarks are much lighter than $b$, namely, $m_b^2/m_c^2\sim 11$, traces of the $cq$ attraction in the color anti-triplet spin-0 state may or may not be present in the $cqq$ baryons.
}

\vspace{1mm}
\begin{center}
{\large\it In Memoriam of James Bjorken (1934-2024)}
\end{center}
\begin{figure}[h]
\begin{center}
\includegraphics[width=4cm]{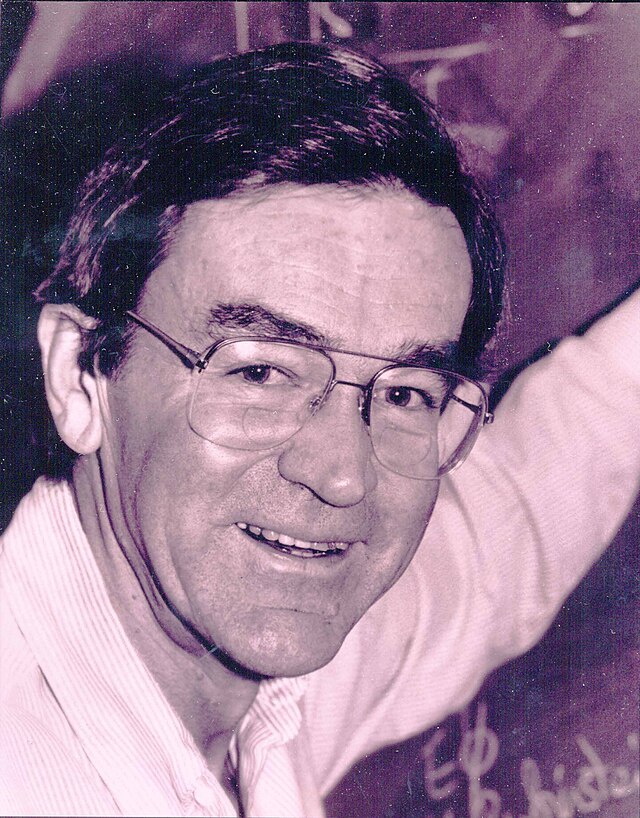}
\end{center}
\vspace{-5mm}
\end{figure} 

\newpage

Although it was not recognized as such, QCD Chemistry exists for almost 50 years, beginning from the mid-1970s. Today I want to discuss 
only one of its aspects, namely, diquarks of the type $Qq$ where $Q$ is a heavy quark ($c$ or $b$) while 
$q$ is a light quark ($u,\, d$ or $s$). I would like to remind what object could  be reasonably described as ``diquark'' and why there is no place for  the $Qq$ diquarks  in QCD in the limit of large $m_Q$. 

\vspace{2mm}

{\em I will start with a brief history.}

\vspace{1mm}

Probably the first QCD ``chemistry'' work was authored by Voloshin and Okun \cite{VO} who argued that molecule-like  states of the $D\bar D $ type exist in the spectrum of the hidden-charm mesons -- the so-called molecular charmonium. In some sense the molecular charmonium is conceptually similar to  quasinuclear resonances \cite{Shap}. Exotic $Q\bar{Q}$-gluon states were discussed approximately at the same time in \cite{OV}.

Quantum chemistry is known from the 1930s. It is based on the solution of the Schr\"odinger equations (with corrections from the Dirac equations where necessary) typically for the multielectron configurations. These configurations  create a complex structure of
inter-electron and electron-nuclei interactions, 
so contrived in large molecules that even numerical calculations on modern computers  are not always sufficient. What is important, is the fact that
at an elementary level the interaction is weak unless the atomic number becomes $\sim 100$ because the corresponding dynamics are largely due to electromagnetic interactions. Pair creation plays no role 
until one approaches a critical value of the nucleus electric charge.

In QCD chemistry one has to deal with strong coupling from the very beginning -- in soft processes the $\alpha_s$ coupling is $\alpha_s\sim 1$, and relativistic effects may be (are) large. Moreover, QCD does {\em not} reduce to any equation of the Schr\"odinger type.  This produces
complexity even in systems containing $\leq$4-5 building blocks or so. This is a typical number of ``players''   under considerations in baryons, tetra- and penta-quarks. If for some reasons a couple of the building blocks attract each other and form a more compact object\,\footnote{One can call it a strong positive correlation.}  than the hadronic state under 
consideration  
then one can view these objects as ``diquarks.'' 
Light diquarks of the type $qq^\prime$ where $q, \, q^\prime = u,\, d$ or $s$ belong to the class of ``good'' diquarks 
provided they form a color-antitriplet spin-singlet pair with antisymmetrization over the light-quark flavors ($[ud], [us]$ or $[ds]$).

Where these quantum numbers come from? In the compact $q, \, q^\prime$ states angular-momentum excitations should be noticeably  heavier that the $L=0$ states. In what follows we will ignore them. Moreover, in SU(3)$_{\rm gauge}$ QCD two quarks in the fundamental (triplet) representation
could form either a color anti-triplet or a sextet. Color-wise the anti-triplet acts as an antiquark which is important for the hadronic architechture. Thus,  the structure we start from is the $\big(q^\alpha_{\,i}\big)^f\big(q^{\,\prime}_{\alpha j}\big)^g\,\varepsilon^{ijk}$. Since  $q, \, q^{\,\prime}$ are anticomuting fermions
the light-quark flavors must be automatically antisymmetrized. This leads us to a good diquark 
$$[q, \, q^\prime]\equiv \big(q^\alpha_{\,i}\big)^{[f}\big(q^{\,\prime}_{\alpha j}\big)^{g]}\,\varepsilon^{ijk}.$$
The square brackets denote antisymetrization. If, instead, we symmetrize over color (marked by braces) leading to the color sextet we will arrive at the bad diquark,
$$\{q, \, q^\prime \}\equiv \big(q^\alpha_{{\color{red}\{ }i}\big)^{\{f}\big(q^{\,\prime}_{\alpha\,j\color{red} \}}\big)^{g\}} .$$
Symmetrization over color mandates symmetrization over flavor.

Early mentions of ``diquarks'' predate QCD \cite{pred}. It is curious that J. Bjorken started working diquarks in the early 1974 and in October 1974 even drafted a paper titled 
``The Diquark and Its Role in Hadron Structure'' \cite{BJ}. A copy of this draft was sent to me by Marek Karliner. Then came the November-74 revolution,
and Bjorken's project was postponed -- indefinitely.
\vspace{2mm}

{\em XXI century}

\vspace{1mm}

In connection with proliferation of experimental data on heavy tetra- and penta-quarks, quite abundant in the last decade (see e.g. \cite{bramb}),
theoretical speculations on heavy-light diquarks of the type $Qq$ became common, see e.g. \cite{data}. The $Qq$ diquarks are the main focus of my review to which I will return at the end. 

One should understand that the notion of diquarks is rather nebulous, even more so than the notion of a constituent quark. Both are {\em not} color-singlets  and do not exist as asymptotic states. Both play no role in hard processes in which at least quarks (but not constituent quarks) are observed.

At the same time, both may be viewed as convenient ``effective objects'' in modeling soft hadronic dynamics, in particular, in spectroscopy. They may be considered for qualitative or semi-quantitive purposes. Information on their masses and other static characteristics can be obtained only indirectly and only approximately. I think, one should not expect 
to attach to these numbers  honest-to-god ``scientific'' error bars.

\vspace{1mm}

The first mention of a special role of diquarks $[su]$ and $[du]$ in $\Delta T=\tfrac 1 2 $ weak decays $\Sigma^+\to p\pi^0$ can be found in \cite{SVZweak}.
The diquark ``density'' introduced there determines the matrix element of the leading operator $O_1$.  See also \cite{stech,dosch}.

In 1981 a classification of hadrons was proposed \cite{NSVZ} which reflects the strength of interaction of various currents with vacuum fields. It was found that the mass and size scales intrinsic to the meson resonances in different channels is not universal; a large scale was discovered in the $0^\pm$  quark and gluon channels with the vacuum quantum numbers. Basing on our observation we established a significant hierarchy of scales. 
In ``normal channels'' such as, say $\rho$ meson, a number of regularities are found 
in hadronic phenomenology and spectroscopy such the Zweig's rule \cite{zweig} also known as the Okubo-Zweig-Iizuka rule, approximate SU(6) symmetry, relatively small values of $\Gamma/m$, and so on. These regularities are badly violated in the light-quark mesons or glueballs of $J^P= 0^\pm$ provided 
that the quark mesons carry only trivial overall flavor numbers. 
  An explanation suggested in \cite{NSVZ}  implied
 that the failure of the above regularities is due to a relatively tight gluon core inside, characterized by unusually strong interaction with the light quark.
 The existence of the core could be related to instantons.
 Further analysis in the framework of a multi-instanton vacuum model was performed in \cite{SSV}.
The study carried out in \cite{SSV} was based on the existing knowledge of the QCD vacuum, in particular, instanton-induced effects. The conclusions was as follows: the  interaction responsible for $0^\pm$ glueballs and quark mesons which can mix with them is stronger than one could have expected {\em a priori}. 

The revival of interest in diquarks in hadronic spectroscopy refers to 2004, see Refs. \cite{JW,jaffe,Wi,SeWi}.
The overall picture that emerged in the 1980s and 1990s  matched the conclusions of Jaffe, Wilczek and Selem\&Wilczek \cite{JW,jaffe,Wi,SeWi} who singled out the spin-0 flavor-antisymmetric diquarks and called them ``good diqurks'', as opposed to flavor-symmetric diquark states, referred to as ``bad diquarks.'' Good diquarks are in  the color-antitriplet representation. Dynamically, in its interactions with gluons, such a diquark behaves as an antiquark.

Estimates for the constituent quark mass are scattered in the interval $(310-360)$ MeV (e.g. \cite{KR}). Almost all authors agree that
the difference between the bad and good diquark masses is $M(\{qq\})- M([qq])\approx$  200 MeV in the $ud$ case. This number can be presented as
follows: 200 = 150+50, the former is associated with the quark attraction due to spin interaction in  $[qq]$ while the latter comes  from the quark 
repulsion in $\{qq\}$. At the same time, estimates of the good $[qq]$ diquark mass 
{\em per se} are less certain. On one hand,  Selem and Wilczek write \cite{SeWi}\label{SWquo}

\begin{quote}
\small
[In Regge trajectories], $L = 2$ appears to be large enough for simple dynamics to apply, and yet has enough data to make the case powerfully. [...]

The near-degeneracy between the baryon and meson states [$N(1680)$  vs. $\rho (1690)$], and between the $\Lambda (1820)$ and $K$ mesons states ... exhibits the near-degeneracy between the good diquark and a light antiquark.
\end{quote}
\label{SWquo2}
The above statement presumably means that the  $[qq]$ mass is less than 400 MeV. On the other hand, according  to \cite{KR}, $M([qq])\sim 560$ MeV. Overall, it seems likely that
$M([qq])$ lies in the interval $(350-550)$ MeV.

\vspace{3mm}
{\em ``Good'' and ``bad'' $qq$ diquarks}

\vspace{1mm}

 The papers by Jaffe \cite{jaffe} and Selem \& Wilczek  \cite{SeWi} were  further developed in \cite{SV}, see also \cite{close,karl,gogo,tamar}, etc. Recent study of the role of light $qq$ good diquarks in nucleons can be found in
\cite{SZ}.
In this publication it was argued that in multiquark systems diquarks should be treated  with care,
and their very existence should not be taken for granted because multiquark interactions could ruin a rather modest attraction in the good diquark channels, see e.g. \cite{MS}. 
\label{p5}

Now, I'd like to review a few extra details of the Selem-Wilczek paper \cite{SeWi}. The authors note that $M(\Sigma_c)-M (\Lambda_c)\approx\, {\rm  170 \,\, MeV}$ implying that the spin interaction of the $c$
is not yet small enough, which is also confirmed by $J/\psi$-$\eta_c$ mass splitting, 
$M(J/\psi)-M (\eta_c)\approx\, {\rm  110 \,\, MeV}$. Since 
\beq
m_b/m_c \approx 3.29\,, 
\eeq
in the $b$-quark containing hadrons the spin interactions are expected
to be considerably weaker. If we assume that the $[ud]$ diquarks act as a light antiquark we can infer the spin interaction
by comparing the following mass differences
\beq
 \delta_B \equiv M(B^*) - M(B)\approx 46\,{\rm MeV}\,,\qquad  \delta_D \equiv M(D^*) - M(D)\approx 167,\,{\rm MeV}
\eeq
with
\beq
\frac{\delta_D}{\delta_B} \approx 3.06\,,
\label{2}
\eeq
which is quite close to (1).

\begin{figure}[h]
\begin{center}
\includegraphics[width=6cm]{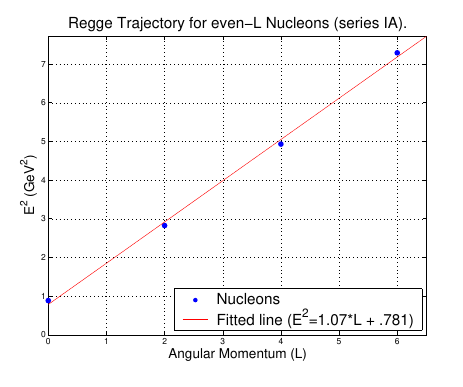}
\end{center}
\vspace{-5mm}
\caption{Nuclear Regge trajectories, from \cite{SeWi}.}
\label{0.0}
\end{figure}

Then  the authors of \cite{SeWi} ignore the spin effects due to $b$-quark and take into account only the $ud$ diquarks, 
spin-1 $\{ud\}$
 in  $\Sigma_b$ and spin-0 $[ud]$ in $\Lambda_b$. The first one is  bad, the second good.
The authors conclude that 
\beq
M(\{qq\})- M([qq])\approx M(\Sigma_b)-M(\Lambda_b) \approx 190\,{\rm MeV}.
\label{THREE}
\eeq
Another estimate of the split (\ref{THREE}) was obtained in \cite{SeWi} by virtue of analyzing the nucleon Regge trajectories (see e.g. Fig. \ref{0.0}) which 
implied
\beq
{\rm Regge \,\, traj}: \qquad M(\{ud\})-M([ud])\sim 240\,\, {\rm Mev}\,.
\label{3}
\eeq
Selem and Wilczek discuss the possibility that the {\em spin attraction effect  in the 
$[ud]$ diquark mass 
is rather close to eating up the mass of one constituent  quarks}, so that $M([ud])\sim M(\bar u)$. At first site
this ``extreme'' hypothesis is in contradiction with a rather large mass difference
\beq
M(\Lambda_b)- M(B) \approx 340\,{\rm MeV}.
\label{laB}
\eeq
However, Selem and Wilczek argue that the above comparison should be carried out not in $L=0$ states as in (\ref{laB})   but rather at large $L$, see the quotation from \cite{SeWi} on page \pageref{SWquo}.
To back up this conjecture they note 
that at  large $L$ there is a marked convergence between the appropriate meson and partner baryon masses due to 
near-equality between $M([ud])$  and $M_{\bar u}$. They also suggest that $L=2$ is large enough and then compare the data with their hypothesis 
 $M([ud])\sim M(\bar u)$. This comparison is quite successful (see the same quotation  on page on page \pageref{SWquo2}). 

Now I will summarize some of the ideas presented in Ref. \cite{SV}.
The usefulness of the diquark notion in hadronic physics is
based on the assumption of an ``abnormally strong'' attraction
in the $0^\pm$ quark and gluon  channels (see observations in \cite{NSVZ}). It is meaningful only
if there are two scales separated by a rather large numerical
factor: the size of a gluon core $R_{\rm core}$ both in  good diquarks and, say, pions compared to a typical
meson size $R_{\rm hadr}$, say, that of the $\rho$ meson. The ratio $R_{\rm hadr}/R_{\rm core}$  may be $\sim 5$ (see below).
 Instanton vs. OPE studies in 1980s \cite{NSVZ} as well as  the instanton liquid model \cite{shu} revealed a numerically large scale in the ``vacuum'' channels  with spin-parity $0^\pm$ 
   in which the 't Hooft regularities miserably fail for low-lying mesons (e.g. we have large decay widths, no OZI rule,       
   etc.) In \cite {SV} we argued that these are the same  short-range correlations that reveal themselves in good diquarks  and are stronger than expected.
   
   The instanton liquid model \cite{shu} provides a particular way of interpreting a hierarchy of scales. The model operates with two parameters, the average instanton size $\rho = 0.48 \Lambda^{-1}$ which is a factor of $\sim 3$ smaller than the average instanton separation QCD
$R = 1.35 \Lambda^{-1}$. Instantons are Euclidean objects and to relate their parameters with
our Minkowski world, note that $R^{-1}\sim \Lambda_{\rm QCD}$ is a typical hadronic scale while $\rho^{-1}$ is the geometrical mean between $\Lambda_{\rm QCD}$  and the higher glueball scale $\Lambda_{\rm gl}$. One can readily verify this at weak coupling in gauge theory with the adjoint Higgs field, where the inverse instanton size is the geometrical mean between the $W$ boson mass and that of monopoles (or spalerons for the Higgs field in the fundamental representation). Hence, we conclude that in the glueball world $\Lambda_{\rm gl} \sim 3^2 \Lambda_{\rm QCD}$, which explains $({\tfrac 1 3})^2$ in Fig. 1.

\vspace{3mm}
{\em Diquarks in {\rm SU(2)$_{\rm gauge}$} QCD}

\vspace{1mm}
   
\begin{figure}[h]
\begin{center}
\includegraphics[width=6cm]{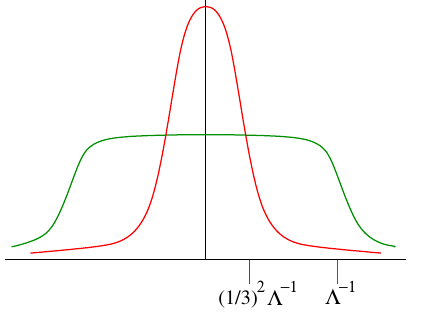}
\end{center}
\vspace{-5mm}
\caption{Small-size cores inside $0^\pm$ mesons with vacuum quantum numbers and good diquarks (red line). The origin of $(\tfrac 1 3)^2$ is explained in \cite{SV} from where this picture is taken.}
\label{00.0}
\end{figure}

   A {\em gedanken} experiment pointing in this direction is as follows.
   Let us replace for a short while ${\rm SU(3)}_{\rm color} \to {\rm SU(2)}_{\rm color}$. Then the $[ud]$ diquarqs become well defined mesons. With two light quarks and ${\rm SU(2)}_{\rm color}$, the $\chi$SB pattern is  ${\rm SU(4)} \to {\rm SO(5)}$. In other words,  in this case we have 5 rather than 3 Goldstones, 3 pions+3 ``diquarks.'' The partnership of pions and good diquarks in
SU(2)$_{\rm color}$ was also discussed in Ref. \cite{dia,RSSV}. In the meson sector the instanton liquid applies both to ${\rm SU(3)}_{\rm color}$ and ${\rm SU(2)}_{\rm color} $. The fact that a typical instanton size is a factor of 3 smaller than the inter-instanton distance 
   leads to the conclusion that pions  have a small-size core in their structure, see Fig. \ref{00.0}. Since in the SU(2)$_{\rm color}$ theory baryonic Goldstones (would-be diquarks in SU(3)$_{\rm color}$) and
pions are related by an exact symmetry, the spatial structure of diquarks  
is the same. This is very important for my final conclusion (see below).
In bona fide QCD with SU(3)$_{\rm color}$, in the intermediate energy interval 
good diquarks act as pointlike color-antitriplet objects whose
interaction with gluons is determined only by the color representation to which
they belong, much in the same way as color-triplet (anti)quarks. This is in full agreement with the instanton liquid model \cite{shuryak2nd}.

\newpage
{\em Weak decays of $b,c$ containing mesons and baryons. }
\vspace{2mm}

The OPE-based weak decay theory for the $b,c$-containing hadrons was worked out in \cite{mimi}.\footnote{Misha Voloshin and I figured out how to estimate
the Pauli interference  -- one of the pre-asymptotic effects -- through the four-fermion
operators and designed relevant
graphs at ITEP canteen in 1982. We actually made
an estimate on a napkin, I put it in a review
with Khoze, and forgot about this, since at
this time I was heavily engaged with SUSY.
A few months later, in 1983, Branko
Guberina and Neven Bilic from Croatia saw it
and detected a wrong sign. They called me. Any call from abroad was highly unusual at that time. To answer the call I was summoned to ITEP International Department. The call was very inspiring. We found the lost sign and decided to resume working on this project.}

For recent reviews see \cite{AL,mis} and original papers
\cite{BL,BL2}. I will returm to \cite{BL} later.

\begin{figure}[h]
\begin{center}
\includegraphics[width=6cm]{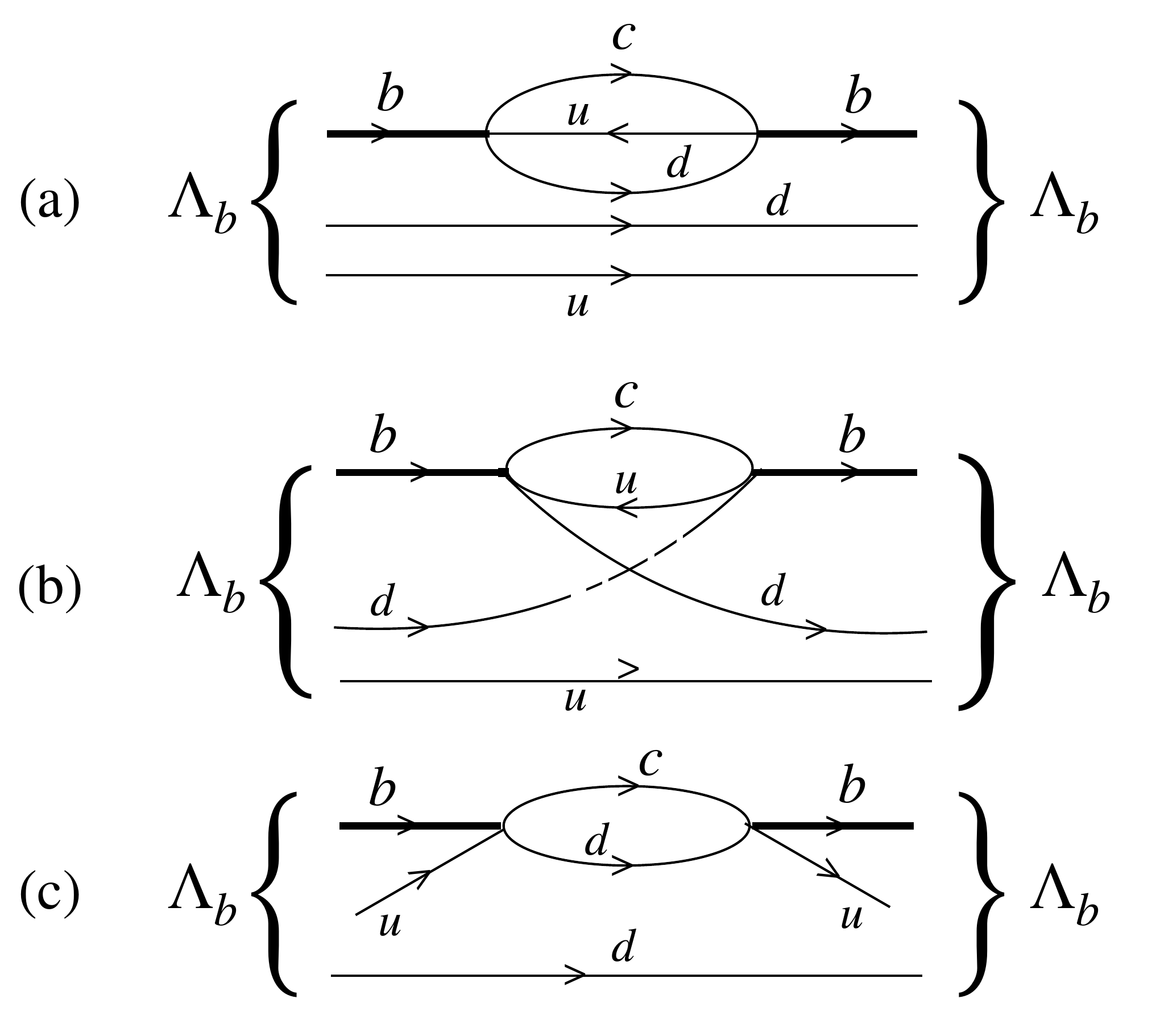}
\end{center}
\caption{\small OPE expansion in powers of  $m_b^{-1}$ for the $\Lambda_b$ lifetime. \\
\hspace{11.99mm} Graph (a) presents the leading terms described by the free heavy quark \\
\hspace{10.59mm}\, decay. Graph (b) describes the $O(m_b^{-3})$ correction to the leading term \\
\hspace{11mm} due to the Pauli interference. Graph (c)  the $O(m_b^{-3})$ correction to the\\
 \hspace{10.2mm} leading term due to the $bu$ ``scattering.'' The latter is 
 proportional to $\langle \Lambda_b |( b u)^\dagger ( b u) | \Lambda_b\rangle$.}\label{last}
\end{figure}

Heavy quarks in QCD introduce  a large scale, $m_Q$.  This is the reason why the hadrons composed from one heavy quark $Q$, a light antiquark $\bar q$, or a good ``diquark'' $qq^\prime$, plus a gluon cloud can be treated in the framework of OPE. The role of the cloud is, of course, to keep all the above objects together, in a colorless bound state. One can visualize the light cloud as a soft medium. The heavy quark $Q$ is then submerged in this medium. The latter circumstance allows one to develop a formalism similar to SVZ in which the soft QCD vacuum medium is replaced by that of the light cloud. As a result, an OPE-based expansion in powers of $1/m_Q$ emerges (see Fig. \ref{last}). In the limit of the large heavy quark mass, the leading term in the width of $H_Q$ is given by the free $Q$ quark decay (Fig. 3a, see also the first term in the second line of Eq. (\ref{WIDTH})). Figures 3b and 3c present preasyptotic OPE corrections in the inverse powers of the heavy quark mass.

A dramatic story of how the theoretical prediction for the ratio $r_{\Lambda_b/B_d}\equiv\tau (\Lambda_b)/ \tau (B_d)$ \cite{bigi} prevailed over the wrong experimental result of the 1990s is narrated in \cite{mis,BL}. The OPE-based theory stated that $1-r_{\Lambda_bB_d}\ <0.1$ (with the most likely prediction around 0.93)  while the early experiments pointed to a much larger deviation.\footnote{At the time we worked on \cite{bigi} the experimental value of $\tau (\Lambda_b)/ \tau (B_d)$ was $0.77\pm0.05$ and our task was to establish what was the largest possible deviation from unity. Giving the benefit of the doubt to negative contributions in our estimates we categorically stated that $1-r_{\Lambda_bB_d}$ larger than 0.1 was impossible.}  What is important for me here is the fact that the 
matrix element    $$\langle \Lambda_b |(  b u)^\dagger (  b u) | \Lambda_b\rangle$$ needed for this calculation (see Fig. \ref{last}c) 
averages the product of two diqaurk operators 
\beq
2 j_k^\dagger\,j_k \,\,\, {\rm where} \, \,\,\boxed{ j_k =\varepsilon_{kji}\,b^j C
\frac{1-\gamma^5}{2} u^i}
\eeq
over $ \Lambda_b$.
If a compact $bu$ diquark existed, this matrix element would be significantly larger than the one used in \cite{bigi, BL}, which, in turn,  will push the OPE-based theory into a dead end.

\begin{center}
\vspace{-0.4cm}
*****
\end{center}
\vspace{-0.4cm}
{\em Exotics with heavy quarks  }
\vspace{1mm}

\begin{figure}[h]
\begin{center}
\includegraphics[width=11cm]{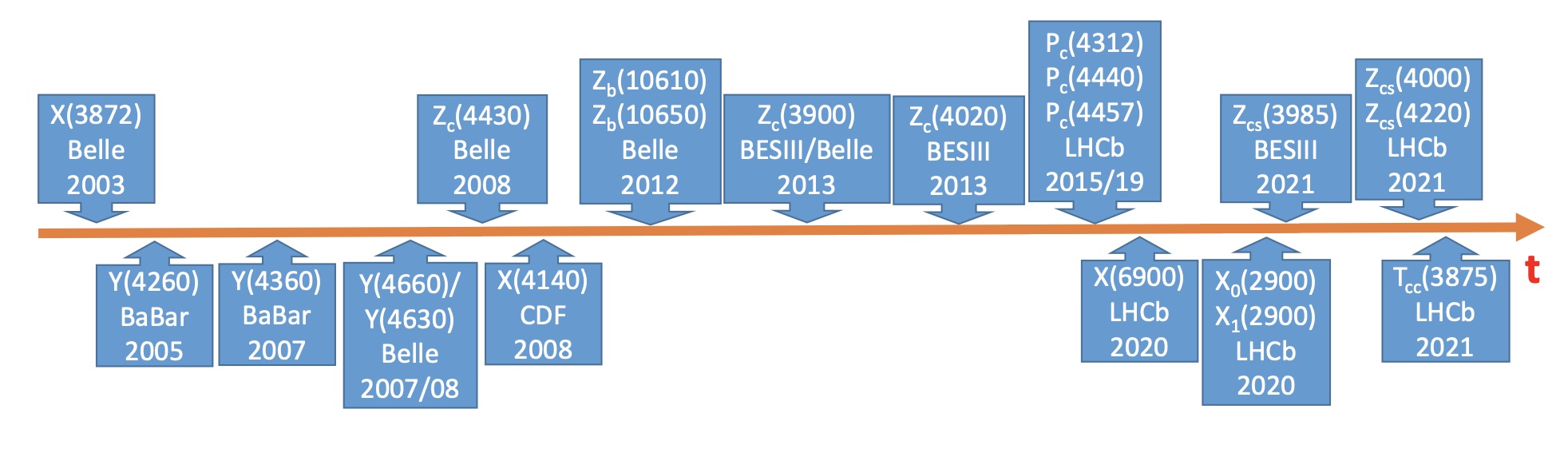}
\end{center}
\vspace{-5mm}
\caption{Discovery of exotic states, from \cite{pos}.}
\label{1.0}
\end{figure}

\begin{figure}[h]
\begin{center}
\includegraphics[width=9cm]{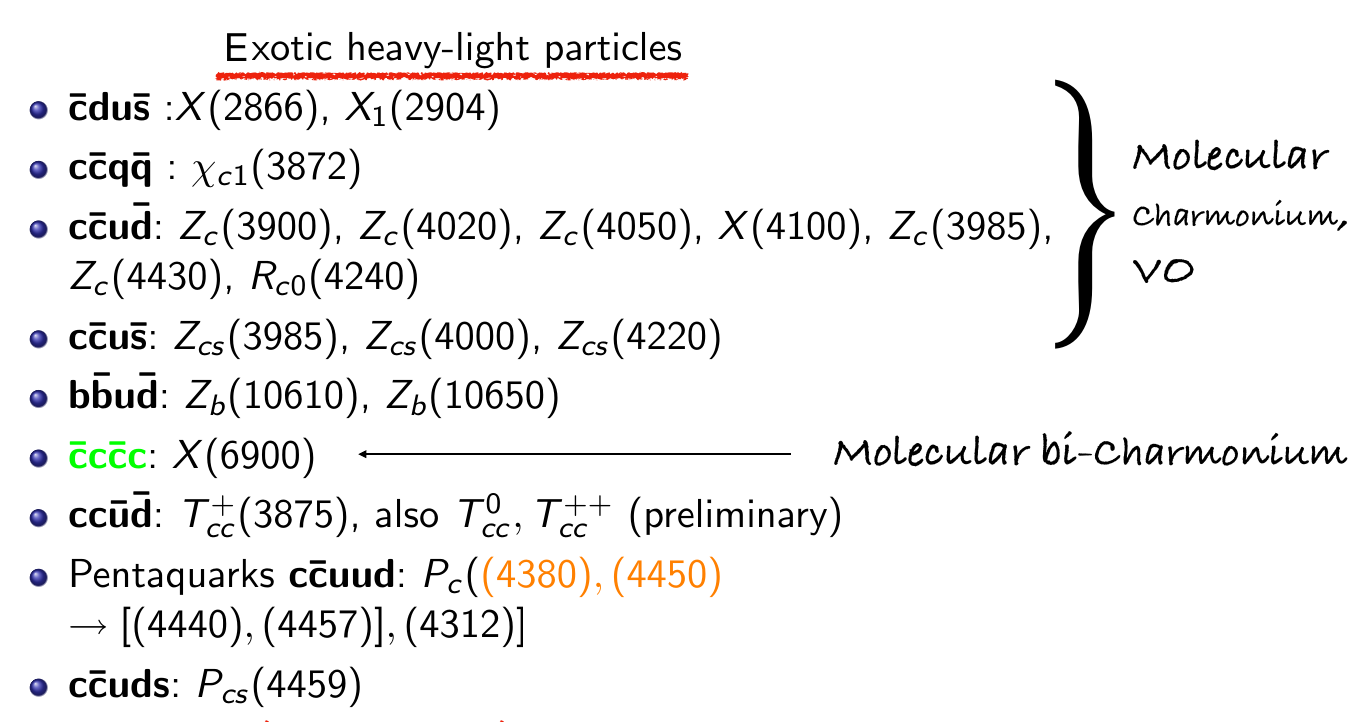}
\end{center}
\vspace{-5mm}
\caption{Quark composition of some exotic states. From Maciej A. Nowak talk at {\sl Baryon 2022}, Seville, Spain. }
\label{2.0}
\end{figure}

Since the beginning of the current century many exotic multiquark states -- tetraquarks, pentaquarks, etc. with heavy quarks  -- have been observed experimentally \cite{pos} (Fig. \ref{1.0}) and discussed theoretically \cite{ali} (Fig. \ref{2.0}). 
 This gave rise to a tsunami wave of diquark studies based on the assumption that $(Qq)$  diquarks play an important role in  the above states.
 

\vspace{2mm}
{\em General considerations  }
\vspace{1mm}

From the previous theoretical studies we know that the very existence of a ``compact good diquark'' is due to a relatively strong attraction
between two quarks in the appropriate channel. Even for the light quarks $u$ and $d$ bad flavor-symmetric diquarks with spin 1 do not form and play no role on the hadronic structure. Spin effects for heavy quarks generally speaking die off as $1/m_Q$. As was mentioned above, even for $c$ quarks they are smaller
compared to Eq. (\ref{3}) (see also page \pageref{p5}); therefore the $cq$ diquarks lie at the borderline and it is unclear on which side. Whether or not the spin attraction in this case is sufficiently potent to form a $[cq]$ quasi-state with a role in hadronic spectroscopy is the question for the future studies.

 For $b$ quarks which can be viewed as {\em bona fide} heavy quarks, the spin attraction in the 
$0^\pm$ channel is smaller still. I expect that for $bu$ the good and bad ``diquarks'' are essentially degenerate and form no compact object as opposed to $du$ good diquark. This is also obvious from the instanton vacuum models. The main conclusion of this talk is that the above statement can be {\em phenomenologically} checked by a thorough analysis 
of weak decays of $b$ containing baryons, more exactly their pre-asymptotic flavor-dependent deviations from the leading order (``parton''-type) results. 
In part, this has been already done in $\Lambda_b$ decays \cite{bigi,BL}. 

\vspace{2mm}

{\em OPE in weak decays of $Qqq$ baryons or how to prove that $Qq$ diquarks do not exist?} 
\vspace{1mm}

The concise narrative on OPE in heavy quarks baryon decays below will be based on \cite{mis,bigi} (Secs. 7-9 and 7, respectively).

One describes the decay rate 
into an 
inclusive final state $f$ in terms of the imaginary part of a  forward
scattering operator (the so-called transition operator)  evaluated to
second order in the  weak interactions \cite{mimi}
\begin{equation} 
{\rm Im}\, \hat T(Q\rightarrow f\rightarrow Q)\;= \;
\, {\rm Im} \int d^4x\ i\,T \left({\cal L}_W(x){\cal 
L}_W^{\dagger}(0)\right)\ 
\label{OPTICAL} 
\end{equation} 
where $T$ denotes the time ordered product and 
${\cal L}_W$ is the relevant weak Lagrangian at the normalization 
point higher or about $m_Q$.  The space-time separation 
$x$ in
Eq.~(\ref{OPTICAL}) is fixed by the inverse energy release. If the 
latter
is 
sufficiently large in the decay, one can express the {\em non-local} 
operator
product in  Eq. (\ref{OPTICAL})  as an infinite sum of {\em local}
operators $O_i$ of increasing  dimensions.  In our case the localization is of the order of $x\sim m_b^{-1}$. 
The width for $H_Q\rightarrow f$ is then  obtained by  averaging  
${\rm Im}\,\hat T$ over the heavy-flavor hadron $H_Q$, which in our example is $\Lambda_b$,
$$
\frac{\langle \Lambda_b| {{\rm Im}\, \hat T (b\to f\to
b)}|{\Lambda_b}}{2M_{\Lambda_b}} \propto 
\Gamma (\Lambda_b\rightarrow f) 
$$
\beq
=G_F^2 |V_{\rm bc}|^2m_b^5   \,
\sum _i  \tilde c_i^{(f)}(\mu ) 
\frac{\langle{\Lambda_b}|{O_i}|{\Lambda_b}\rangle_{\mu }}{2M_{\Lambda_b}}    
\label{OPE}
\end{equation} 
where I neglect the highly suppressed mode $b\to u$.  The parameter $\mu$ in Eq. (\ref{OPE}) is the normalization point, 
indicating that we explicitly evolved from $m_Q$ down to $\mu$. 
The
effects of momenta {\em below}  $\mu$  are presented by the matrix 
elements of the operators $O_i$. 
The coefficients $\tilde c_i^{(f)}(\mu )$ are dimensionful, they 
contain powers of 
$1/m_Q$. We calculated $\tilde c_i^{(f)}(\mu )$ using perturbation theory and assuming quark-hadron duality \cite{qhd} at energies $E\sim m_b$. 
Violations of duality due to nonperturbative effects 
are believed to be rather small in 
beauty decays;  they may be quite significant, however, in the $c$-quark decays.

For non-leptonic decays of $\Lambda_b$ the master formula is
\beqn
\Gamma (\Lambda_b\ra f)&=&\frac{G_F^2m_b^5}{192\pi ^3}|V_{\rm 
bc}|^2
\nonumber\\[2mm]
&\times&
\left[ c_3(\mu )\frac{\langle\Lambda_b|\bar{b}b|\Lambda_b\rangle_\mu}{2M_{\Lambda_b}}
\right.
+ c_5(\mu ) m_b^{-2}
\frac{
\matel{\Lambda_b}{\bar b\frac{i}{2}\sigma G b}{\Lambda_b}_{\mu}}
{2M_{\Lambda_b}} 
\nonumber\\[2mm]
 &+& \sum _i c_{6,i}(\mu )m_b^{-3}
 \frac{\langle \Lambda_b | (\bar b\Gamma _iq)(\bar q\Gamma _ib)|\Lambda_b\rangle_\mu}{2M_{\Lambda_b}}
\nonumber\\[2mm]
&+&\left. c_{6}(\mu )m_b^{-3} \frac{\langle \Lambda_b |( b u)^\dagger ( b u) | \Lambda_b\rangle_\mu}{2M_{\Lambda_b}}+ {\cal O}(1/m_b^4)\right]  \, .
\label{WIDTH} 
\eeqn
Here $c_{3,5,6}$ are the OPE coeffcients, the subscripts 3,5,6 indicate the dimensions of the corresponding operators, $\mu$ is the normalization point assumed to be of the order of 
$\Lambda_{\rm QCD}$, and the subscript $i$ marks various combinations of the Dirac matrices $\gamma$. The four-fermion operator in the last line 
plays a special role. It emerges from the graph $3c$ describing the $bu$ ``scattering'' mechanism in preasymptotic  corrections which 
does not exist for $b\bar q$ mesons. Numerically it is larger than the four-fermion operator on the third line.
Let us consider it in more detail with regards to the possible role of the $bu$ diquark. 

In Refs. \cite{BL,BL2} a non-relativistic quark model (NQM) is used to average the four-fermion operators over the baryon states of the type of $\Lambda_b$.
Roughly speaking, in the $S$ wave  the ratio
\beq
\frac{\langle \Lambda_b |( b u)^\dagger ( b u) | \Lambda_b\rangle}{\langle\Lambda_b|\bar{b}b|\Lambda_b\rangle}\sim \frac{1}{r^3}
\eeq
where $r$ is the ``radius'' of the $bu$ cloud. If $bu$ is relatively tightly bound, as expected for a diquark, then $r\sim r_{\rm dq}$. If, however, there is no special correlation between $b$ and $u$ in the color anti-triplet $S$-wave state then $r\sim  r_{\Lambda_b}$. In NQM we have the latter situation. 
Since $r_{\Lambda_b}$ is expected to be significantly larger than $r_{\rm dq}$ (see Fig. 2) it is natural to think within the diquark assumption the predictions for the preasymptotic
correction in $\Gamma (\Lambda_b\ra f)$ will be significantly larger than in \cite{BL,BL2} which would contradict experimental data. Thus we will be able to conclude that the heavy-light diquarks ($bu$ in the case of $\Lambda_b$)   do not form in $bqq$ baryons.

\vspace{2mm}
{\em Acknowledgments}

\vspace{1mm}

I am grateful to S. Narison for his kind invitation to deliver a talk at the 27$^{\rm th}$ High-Energy Physics International Conference in Quantum Chromodynamis (QCD24), Montpellier, July  8-12, 2024. I would like to say thank you to Marek Karliner, Alexander Lenz, Bla\v{z}enka Meli\'c, Edward Shuryak and Arkady Vainshtein for many useful discussions. 

This work is supported in part by DOE grant DE-SC0011842 and the Simons Foundation Targeted Grant 920184 to the Fine Theoretical Physics Institute.

\vspace{4mm}


\vspace{1cm}

\begin{thebibliography}{99}

{\small

\bibitem{VO} 
M.~B.~Voloshin and L.~B.~Okun,
{\em Hadron Molecules and Charmonium Atom,}
JETP Lett. \textbf{23}, 333-336 (1976).

\bibitem{Shap}
I.S. Shapiro, {\em The Physics of Nucleon -- anti-Nucleon Systems,}
Phys. Rept. \textbf{35}, 129-185 (1978); {\em Some New Features in Low-energy Anti-proton Physics,}
Nucl. Phys. A \textbf{478}, 665C-672C (1988) and a representative list of references therein.

\bibitem{OV}
A.I. Vainshtein,  L.B. Okun,
{\em Anomalous Levels of Charmonium}, Yad. Fiz. \textbf{23}, 1347-1348, (1976).

\bibitem{pred}
M.~Ida and R.~Kobayashi,
{\em Baryon resonances in a quark model,}
Prog. Theor. Phys. \textbf{36}, 846 (1966);
%doi:10.1143/PTP.36.846
D.~B.~Lichtenberg and L.~J.~Tassie,
{\em Baryon Mass Splitting in a Boson-Fermion Model,}
Phys. Rev. \textbf{155}, 1601-1606 (1967)
%doi:10.1103/PhysRev.155.1601

\bibitem{BJ} 
James Bjorken, {\em The Diquark and Its Role in Hadron Structure}, 1974, unpublished; courtesy of Marek Karliner.

\bibitem{bramb}
N.~Brambilla, H.~X.~Chen, A.~Esposito, J.~Ferretti, A.~Francis, F.~K.~Guo, C.~Hanhart, A.~Hosaka, R.~L.~Jaffe and M.~Karliner, \textit{et al.}
{\em Substructure of Multiquark Hadrons} (Snowmass 2021 White Paper),'
[arXiv:2203.16583 [hep-ph]].
%22 citations counted in INSPIRE as of 27 Nov 2024

\bibitem{data}
J.~Q.~Xie, H.~Song, X.~Feng and J.~K.~Chen,
{\em $\ensuremath{\lambda}$ and $\ensuremath{\rho}$ Regge trajectories for hidden bottom and charm tetraquarks $(Qq)(Qq)$},
Phys. Rev. D \textbf{110}, no.7, 074039 (2024)
%doi:10.1103/PhysRevD.110.074039
[arXiv:2407.04222 [hep-ph]];

J.~K.~Chen, X.~Feng and J.~Q.~Xie,
{\em Regge trajectories for the heavy-light diquarks,}
JHEP \textbf{10}, 052 (2023)
%doi:10.1007/JHEP10(2023)052
[arXiv:2308.02289 [hep-ph]].

T.~de Oliveira, D.~Harnett, R.~Kleiv, A.~Palameta and T.~G.~Steele,
{\em ``Light-Quark $SU(3)$ Flavour Splitting of Heavy-Light Constituent Diquark Masses and Doubly-Strange Diquarks from QCD Sum-Rules,}
Phys. Rev. D \textbf{108}, no.5, 054036 (2023)
%doi:10.1103/PhysRevD.108.054036
[arXiv:2307.15815 [hep-ph]].

O.~Andreev,
{\em $QQqqq$ quark system, compact pentaquark, and gauge/string duality. II,}
Phys. Rev. D \textbf{108}, no.10, 106012 (2023)
%doi:10.1103/PhysRevD.108.106012
[arXiv:2306.08581 [hep-ph]].

R.~Tiwari, J.~Oudichhya and A.~K.~Rai,
{\em Mass-spectra of light-heavy tetraquarks,}
Int. J. Mod. Phys. A \textbf{38}, no.33n34, 2341007 (2023)
%doi:10.1142/S0217751X23410075
[arXiv:2205.00679 [hep-ph]].

R.~Tiwari, D.~P.~Rathaud and A.~K.~Rai,
{\em Mass-spectroscopy of} [$bb][{\bar{b}}{\bar{b}}$] {\rm and} [$bq][{\bar{b}}{\bar{q}}$] {\rm tetraquark states in a diquark\textendash{}antidiquark formalism,}
Eur. Phys. J. A \textbf{57}, no.10, 289 (2021)
%doi:10.1140/epja/s10050-021-00601-w
[arXiv:2108.06521 [hep-ph]].

\bibitem{SVZweak}
M.~A.~Shifman, A.~I.~Vainshtein and V.~I.~Zakharov,
{\em  Nonleptonic Decays of K Mesons and Hyperons,}
Sov. Phys. JETP \textbf{45}, 670 (1977), Sec. 5.

\bibitem{stech}
B. Stech, {\em $|\Delta I|= \tfrac 12 $
 rule and consequences for $D$ and $B$ decays and $\frac{\varepsilon^\prime}{\varepsilon}$.}\\
Phys. Rev. D 36, 975 (1987).

\bibitem{dosch}
H. G. Dosch, M. Jamin, and B. Stech, {\em Diquarks, QCD sum rules, and weak decays},
Z. Phys. C 42, 167 (1989).

\bibitem{NSVZ}
V.~A.~Novikov, M.~A.~Shifman, A.~I.~Vainshtein and V.~I.~Zakharov,
{\em Are All Hadrons Alike?},
Nucl. Phys. B \textbf{191}, 301-369 (1981).

\bibitem{zweig}
 G. Zweig, CERN-TH Report 412 ``An SU$_3$
 model for strong interaction symmetry and its breaking,'' (1964);\\
S. Okubo, $\phi${\em -meson and unitary symmetry model}, Phys. Lett. \textbf{5}, 165-168 (1963);\\
J. Iizuka, {\em Systematics and Phenomenology of Meson Family}, Prog. Theor. Phys. Suppl. \textbf{37}, 21-34 (1966).

\bibitem{SSV}
T.~Sch\"afer, E.~V.~Shuryak and J.~J.~M.~Verbaarschot,
{\em Baryonic correlators in the random instanton vacuum,}
Nucl. Phys. B \textbf{412}, 143-168 (1994)
%doi:10.1016/0550-3213(94)90497-9
[arXiv:hep-ph/9306220 [hep-ph]].

\bibitem{JW}
R.~L.~Jaffe and F.~Wilczek,
{\em Diquarks and exotic spectroscopy,}
Phys. Rev. Lett. \textbf{91}, 232003 (2003)
%doi:10.1103/PhysRevLett.91.232003
[arXiv:hep-ph/0307341 [hep-ph]].

\bibitem{jaffe}
R.~L.~Jaffe,
{\em Exotica,}
Phys. Rept. \textbf{409}, 1-45 (2005)
%doi:10.1016/j.physrep.2004.11.005
[arXiv:hep-ph/0409065 [hep-ph]].

\bibitem{KR}
M. Karliner and J. Rosner,
{\em Baryons with two heavy quarks: Masses, production, decays, and detection}'',
Phys. Rev. {\bf D 90} (2014), 094007, Section II,
%doi:10.1103/PhysRevD.90.094007
[arXiv:1408.5877 [hep-ph]].

\bibitem{Wi}
F.~Wilczek,
{\em Diquarks as inspiration and as objects,}
[arXiv:hep-ph/0409168 [hep-ph]]; Published in: {\sl From Fields to Strings: Circumnavigating Theoretical Physics}, 
Ian Kogan Memorial Collection, Ed. M. Shifman et al., Vol. 1, page 77 (World Scientific, Singapore, 2005).

\bibitem{SeWi}
A.~Selem and F.~Wilczek,
{\em Hadron systematics and emergent diquarks,}
%doi:10.1142/9789812773524\_0030
[arXiv:hep-ph/0602128 [hep-ph]];  published in Proc. Ringberg Workshop on New Trends in HERA Physics 2005
(2-7 October 2005. Ringberg Castle, Tegernsee, Germany), World Scientific, Singapore, pages  337-356.



\bibitem{SV}
M. Shifman, A. Vainshtein
{\em Remarks on diquarks, strong binding, and a large hidden QCD scale},
Phys. Rev. D \textbf{71}, 074010 (2005)
%doi:10.1103/PhysRevD.71.074010
[arXiv:hep-ph/0501200 [hep-ph]].

\bibitem{close}
F.~E.~Close,
{\em Quarks, diquarks, tetraquarks, and pentaquarks,}
Contemp. Phys. \textbf{47}, 67-78 (2006)

\bibitem{karl}
M.~Karliner and H.~J.~Lipkin,
{\em Diquarks and antiquarks in exotics: A Menage a trois and a menage a quatre,}
Phys. Lett. B \textbf{638}, 221-228 (2006)
%doi:10.1016/j.physletb.2006.05.032
[arXiv:hep-ph/0601193 [hep-ph]].

\bibitem{gogo}
I.~Gogoladze, Y.~Mimura, N.~Okada and Q.~Shafi,
{\em Color Triplet Diquarks at the LHC,}
Phys. Lett. B \textbf{686}, 233-238 (2010)
%doi:10.1016/j.physletb.2010.02.068
[arXiv:1001.5260 [hep-ph]].

\bibitem{tamar} 
T.~Friedmann,
{\em No Radial Excitations in Low Energy QCD. I. Diquarks and Classification of Mesons,}
Eur. Phys. J. C \textbf{73}, no.2, 2298 (2013)
doi:10.1140/epjc/s10052-013-2298-9
[arXiv:0910.2229 [hep-ph]].

\bibitem{SZ}
E.~Shuryak and I.~Zahed,
{\em Hadronic structure on the light front. V. Diquarks, nucleons, and multiquark Fock components,}
Phys. Rev. D \textbf{107}, no.3, 034027 (2023)
%doi:10.1103/PhysRevD.107.034027
[arXiv:2208.04428 [hep-ph]].

\bibitem{MS}
N.~Miesch and E.~Shuryak,
{\em Wave functions of multiquark hadrons from representations of the symmetry groups $S_n$,}
[arXiv:2406.05024 [hep-ph]].

\bibitem{shu}
E.V. Shuryak, Nucl. Phys. B 203, 93 (1982); 203, 116
(1982); 203, 140 (1982); 302, 559 (1988); 302, 574
(1988); 302, 599 (1988); 302, 621 (1988); D. Diakonov
and V.Y. Petrov, Nucl. Phys. B 272, 457 (1986); 245, 259
(1984); for a review, see E.V. Shuryak and T. Sch\"afer,
Annu. Rev. Nucl. Part. Sci. 47, 359 (1997).

\bibitem{dia}
D. Diakonov and V. Petrov, in {\sl Quark Cluster Dynamics},
Lecture Notes in Physics, edited by K. Goeke, P. Kroll,
and H.-R. Petry (Springer-Verlag, Berlin, 1992), p. 288.

\bibitem{RSSV} 
R.~Rapp, T.~Sch\"afer, E.~V.~Shuryak and M.~Velkovsky,
{\em Diquark Bose condensates in high density matter and instantons,}
Phys. Rev. Lett. \textbf{81}, 53-56 (1998)
%doi:10.1103/PhysRevLett.81.53
[arXiv:hep-ph/9711396 [hep-ph]].

\bibitem{shuryak2nd}
E. Shuryak, {\sl QCD Vacuum, Hadrons and  Superdense Matter},  2nd Edition,  (World Scientific, Singapore, 2004). 

\bibitem{mimi}
M. Shifman and M. Voloshin, {\em Preasymptotic Effects in Inclusive Weak Decays of Charmed Particles}, Sov. J. Nucl. Phys. 41, 120 (1985); 
{\em Hierarchy of Lifetimes of Charmed and Beautiful Hadrons}, Sov. Phys. JETP 64, 698 (1986). The first formulation of the idea 
of applying OPE to weak decays of heavy-quark hadrons can be found in V.~A.~Khoze and M.~A.~Shifman,
{\em Heavy Quarks},
Sov. Phys. Usp. \textbf{26}, 387 (1983), Sect. 3d, with reference to Shifman, Voloshin, unpublished.

\bibitem{AL}
Alexander Lenz, {\em Lifetimes and heavy quark expansion}, Int. J. Mod. Phys. A, 30, 1543005 (2015) [arXiv: 1405.3601].

\bibitem{mis}
M. Shifman, {\em OPE-based Sum Rules}, see p. 153-161 in  
F.~Gross, E.~Klempt, S.~J.~Brodsky, A.~J.~Buras, V.~D.~Burkert, G.~Heinrich, K.~Jakobs, C.~A.~Meyer, K.~Orginos and M.~Strickland, \textit{et al.}
{\em 50 Years of Quantum Chromodynamics,}
Eur. Phys. J. C \textbf{83}, 1125 (2023),  
%doi:10.1140/epjc/s10052-023-11949-2
[arXiv:2212.11107 [hep-ph]] and M. Shifman, {\em OPE-based Methods in Nonperturbative QCD,} [arXiv:2208.10600 [hep-ph]].

\bibitem{BL}
J.~Gratrex, A.~Lenz, B.~Meli\'c, I.~Ni\v{s}and\v{z}i\'c, M.~L.~Piscopo and A.~V.~Rusov,
{\em Quark-hadron duality at work: lifetimes of bottom baryons,}
JHEP \textbf{04}, 034 (2023)
%doi:10.1007/JHEP04(2023)034
[arXiv:2301.07698 [hep-ph]] and references therein.

\bibitem{BL2}
D. King, A. Lenz, M. L. Piscopo, T. Rauh, A. V. Rusov, and C. Vlahos, {\em Revisiting inclusive decay widths of charmed mesons}, JHEP 08 (2022) 241, [arXiv:2109.13219];
J. Gratrex, B. Meli\'c, and I. Ni\v{s}and\v{z}i\'c, {\em Lifetimes of singly charmed hadrons}, JHEP 07 (2022) 058, [arXiv:2204.11935];
A. Lenz, M. L. Piscopo, and A. V. Rusov, {\em Disintegration of beauty: a precision study}, JHEP 01 (2023) 004, [arXiv:2208.02643].

\bibitem{bigi}
I.~I.~Y.~Bigi, M.~A.~Shifman and N.~Uraltsev,
{\em Aspects of heavy quark theory,}
Ann. Rev. Nucl. Part. Sci. \textbf{47}, 591-661 (1997)
%doi:10.1146/annurev.nucl.47.1.591
[arXiv:hep-ph/9703290 [hep-ph]].

\bibitem{pos}
C.~Z.~Yuan,
{\em New experimental results on light and heavy hadrons,}
PoS \textbf{PANIC2021}, 016 (2022)
%doi:10.22323/1.380.0016
[arXiv:2111.07514 [hep-ex]].

\bibitem{ali}
A.~Ali, J.~S.~Lange and S.~Stone,
{\em Exotics: Heavy Pentaquarks and Tetraquarks,}
Prog. Part. Nucl. Phys. \textbf{97}, 123-198 (2017)
%doi:10.1016/j.ppnp.2017.08.003
[arXiv:1706.00610 [hep-ph]].

\bibitem{qhd}
M. Shifman,
{\em Quark hadron duality,}
%doi:10.1142/9789812810458\_0032
[arXiv:hep-ph/0009131 [hep-ph]], Czech. J. Phys. \textbf{52}, B102-B135 (2002);
%283 citations counted in INSPIRE as of 03 Nov 2024
{\em Highly excited hadrons in QCD and beyond}, 
[arXiv:hep-ph/0507246 [hep-ph]], published in
{\sl Quark-Hadron Duality and the Transition to PQCD},
Proceedings of the First Workshop, Frascati, Italy, 2005,
Eds. Alessandra Fantoni, Simonetta Liuti, Oscar A Rond\'on, (World Scientific, 2006), pages 171-191.

}
\end{thebibliography}
\end{document}